\documentstyle[12pt]{article}
\newcommand{\be}{\begin{equation}}
\newcommand{\ee}{\end{equation}}
\newcommand{\ba}{\begin{eqnarray}}
\newcommand{\ea}{\end{eqnarray}}

\begin{document}
\begin{center}
{\bf TWO-DIMENSIONAL $SUSY-$PSEUDO-HERMITICITY WITHOUT SEPARATION OF VARIABLES}\\
\vspace{1cm}
{\large \bf F. Cannata $^{1,}$\footnote{E-mail: cannata@bo.infn.it},
M. V. Ioffe $^{2,}$\footnote{E-mail: m.ioffe@pobox.spbu.ru}
\footnote{{\it Corresponding author.}
Phone:+7(812)4284553; FAX: +7(812)4287240},
D. N. Nishnianidze} $^{2,3,}$\footnote{E-mail: qutaisi@hotmail.com}\\
\vspace{0.5cm}
$^1$ Dipartimento di Fisica and INFN, Via Irnerio 46, 40126 Bologna, Italy.\\
$^2$ Sankt-Petersburg State University,198504 Sankt-Petersburg, Russia\\
$^3$ Faculty of Physics, Kutaisi Polytechnical University, Kutaisi, Republic of
Georgia
\end{center}
\vspace{0.5cm}
\hspace*{0.5in}
\vspace{1cm}
\hspace*{0.5in}
\begin{minipage}{5.0in}
{\small
We study $SUSY-$intertwining for non-Hermitian Hamiltonians with
special emphasis to the two-dimensional generalized Morse potential,
which does not allow for separation of variables. The complexified
methods of $SUSY-$separation of variables and two-dimensional shape
invariance are used to construct particular solutions - both for
complex conjugated energy pairs and for non-paired complex energies.
}
\end{minipage}



\section*{\normalsize\bf 1. \quad Introduction}
\vspace*{0.5cm}
\hspace*{3ex}
Recently $PT-$invariance of one-dimensional models in Quantum
Mechanics has been investigated by C.Bender and collaborators
\cite{bender0}-\cite{bender02}
(see also \cite{tateo})
with special emphasis on the spectrum of the
associated Hamiltonians. Since in many cases the spectrum was found to be real,
$PT-$invariance was proposed as a generalization of standard Hermiticity.
However it soon became clear that there are simple $PT-$symmetric examples,
for which the spectrum is not real and therefore alternative criteria
for reality of the spectrum were explored.

The most systematic investigation has been performed by
A.Mostafazadeh \cite{most}-\cite{most2}
(see also \cite{ahmed}, \cite{japaridze})
elaborating on the so-called pseudo-Hermiticity:
\be
\eta H \eta^{-1} = H^{\dagger} \label{ps}
\ee
with $\eta$ a Hermitian invertible operator, expressed
in terms of a biorthogonal basis. This type of approach requires
a complete solution of the spectral problem.

For non-solvable problems, it is convenient to use the
intertwining relations of SUSY Quantum Mechanics (SUSY QM)
\cite{review},\cite{review1}
to relate pairs
of Hamiltonians. One Hermitian and one non-Hermitian Hamiltonian
may be intertwined \cite{ACDI2}, \cite{junker}, \cite{fernandez}
or, in other cases, non-Hermitian Hamiltonians may be
intertwined \cite{bagchi2}-\cite{bagchi21}
($SUSY-${\bf pseudo-Hermiticity}) with their Hermitian conjugates.
Both these constructions might lead to complex models with real spectra.
We would like to remark that $SUSY-$pseudo-Hermiticity differs from
the pseudo-supersymmetry of \cite{mostps}.

While one dimensional models of this kind have been
investigated in many different frameworks like SUSY QM \cite{ACDI2},
\cite{junker}, \cite{znojil}, \cite{raj},
$PT$ symmetry \cite{bender0}-\cite{bender02},
\cite{znojilPT}-\cite{znojilPT3}, \cite{raj},
for two dimensions the advance is really at the start. To our
knowledge there are only the papers \cite{bender2} and
\cite{srilanka}, where some
complex two-dimensional potentials are studied numerically,
which are $PT-$symmetric and therefore are
two dimensional extensions of the $x^2+igx^3$ potential \cite{caliceti}.

Within SUSY QM a class of non-trivial two-dimensional
models (not allowing for separation of variables) was constructed in
\cite{david}-\cite{david3}. One model of this class (generalized
singular Morse
potential) was investigated \cite{two} in detail by two novel methods:
$SUSY-${\bf separation of variables} and {\bf two-dimensional shape
invariance}.
The model is partially solvable (see, for example
\cite{finkel})
or, in alternative terminology, quasi-exactly-solvable
\cite{turbiner},\cite{turbiner1},
this means that only part of eigenvalues and eigenfunctions can be found.

In Section 2 we introduce $SUSY-${\bf pseudo-Hermiticity}
with supercharges of first and second order in derivatives in one and two
dimensions. In Section 3 the complexification of two-dimensional model
of \cite{two} is implemented in the context of $SUSY-$pseudo-Hermiticity
with special attention to the appearance of levels in
complex conjugated pairs. In particular, Subsection 3.1 contains
the $SUSY-$separation of variables method, and 3.2 - the
complex form of the two-dimensional shape invariance method.

\section*{\normalsize\bf 2.\quad SUSY QM and $SUSY-$pseudo-Hermiticity.}
\vspace*{0.5cm}
\hspace*{3ex}
For the case of Hermitian Hamiltonians
the main algebraic relation of SUSY Quantum Mechanics
\cite{review},\cite{review1},
in all possible
formulations and generalizations (for example, \cite{ABEI}-\cite{aaa} )
is given by intertwining relations:
\ba
\tilde HQ^+ = Q^+H;  \label{intertw}\\
\quad Q^-\tilde H = HQ^-;\quad  Q^- = (Q^+)^{\dagger} \label{iintertw}
\ea
for a pair of Schr\"odinger operators (superpartners):
$$
\tilde H = -\Delta + \tilde V; \quad H = -\Delta + V.
$$
These relations connect eigenfunctions with the same
eigenvalues (up to zero modes of $Q^{\pm}$):
$$
H \phi_{E_n} = E_n \phi_{E_n};
\quad \tilde H \tilde\phi_{E_n} = E_n \tilde\phi_{E_n};
$$
$$
\phi_{E_n} = Q^- \tilde\phi_{E_n};\quad \tilde\phi_{E_n} = Q^+ \phi_{E_n}.
$$
For compactness we do not introduce explicitly an index associated to
possible degeneracy.

If $H$ and $\tilde H$ are non-Hermitian, the two intertwining relations
(\ref{intertw}) and (\ref{iintertw}) may become independent,
and the supercharges
$Q^{\pm}$ not necessarily Hermitian conjugate. A particular case,
considered earlier \cite{ACDI2}, $\tilde H$ - Hermitian,
and $H$ - not, leads to the reality of the spectrum of $H.$

Another possibility is to examine non-Hermitian partner Hamiltonians
related by what we call {\bf $SUSY-$pseudo-Hermiticity}:
\ba
Q^+H = H^{\dagger}Q^+ ;\label{qh}\\
HQ^- = Q^-H^{\dagger}. \label{hq}
\ea
Since we deal with scalar potentials, Hermitian and complex conjugations
are equivalent for our purposes (see also \cite{aaa}).
The eigenstates of $H^{\dagger}$ with eigenvalues $E_n$ will be
denoted by
$\bigl ( \phi_{E_n^{\star}}\bigr )^{\star}$ with eigenvalues $E_n.$
They are related to those of
$H$ not only by the intertwining but also by direct conjugation.
This has been established by using the formalism of the biorthogonal expansion
\cite{most}-\cite{most2}.
The operator $Q^+$ has in general zero modes becoming non-invertible.

As a consequence of Eq.(\ref{qh}), one can obtain a relation which can be useful
for the classification of the spectrum:
\be
(E_n - E_m^{\star}) <\phi_{E_m}| Q^+\phi_{E_n}> = 0.
\label{spectrum}
\ee
First, let us notice that diagonal matrix elements in the subspace of
zero modes of $Q^+$ are trivially zero. So, in this case (\ref{spectrum})
does not provide any restriction on the energy $E_n ,$ which in
particular can be complex having no complex conjugate partner.
Clearly, a non-zero value of the matrix element in (\ref{spectrum})
for $n=m$ outside the subspace, considered above,
implies that the energy $E_n$ is real, while off-diagonal
non-vanishing matrix element signals that complex energies appear
in complex conjugated pairs $E_n = E_m^{\star}.$

For introducing $SUSY-$pseudo-Hermiticity we provide an explicit
exhaustive construction
in the framework of complex SUSY QM \cite{ACDI2}
in one-dimension with first order supercharges
\cite{bagchi2},\cite{bagchi21}.
In this case without loss of generality:
$$
H=Q^-Q^+ + Const;\quad H^{\dagger}=Q^+Q^- + Const^{\star};
$$
$$
Q^+=-\partial +ig(x);\quad Q^-=+\partial +ig(x)=-(Q^+)^{\star};\quad
\partial\equiv\frac{d}{dx}
$$
with $Const$ an arbitrary complex number and $g(x)$ an arbitrary real function,
leading to the potential:
$$
V(x)=-g^2(x) + ig^{\prime}(x) + Const.
$$
In this case the equation for zero modes of $Q^+$ has no
normalizable solution, so that $SUSY-$pseudo-Hermiticity is effectively
equivalent to pseudo-Hermiticity (\ref{ps}). Connection with usual
$PT-$symmetry can be established by choosing $g(x)$ to be of even parity.
Second order generalizations (HSUSY) of one-dimensional supercharges
\cite{ACDI2} for $SUSY-$pseudo-Hermiticity were recently
discussed \cite{bagchi2},\cite{bagchi21}.

For two-dimensional SUSY QM models solutions of the intertwining
relations (\ref{intertw}) for scalar Schr\"odinger
Hamiltonians only exist with second order supercharges
\cite{david}-\cite{two} and only particular solutions
have been found  by suitable ansatzes.
For the supercharges with Lorentz metrics \cite{two}
\be
Q^+ = (\partial_1^2 - \partial_2^2)
+  C_k \partial_k + B = 4\partial_+\partial_- +C_+\partial_- +
C_-\partial_+ + B, \label{ourq}
\ee
a solution of (\ref{intertw}) can be provided \cite{david}-\cite{david3}
by solving the system:
\ba
&&\partial_-(C_- F) =
-\partial_+(C_+ F);\label{first}\\
&&\partial_+^2 F = \partial_-^2 F,\label{second}
\ea
where $x_{\pm} \equiv x_1\pm x_2\quad
\partial_{\pm}\equiv\partial / \partial x_{\pm} $ and $C_{\pm}$ depend only on
$ x_{\pm},$ respectively:
$$C_+ \equiv C_1 - C_2 \equiv C_+(x_+);\quad
C_- \equiv C_1 + C_2 \equiv C_-(x_-).
$$
The function
$ F, $ solution of (\ref{second}), is represented as
$$ F=F_{1}(x_{+}+x_{-}) + F_{2}(x_{+}-x_{-}).$$
The potentials $ \tilde V(\vec x), V(\vec x) $ and the function $ B(\vec x) $
are expressed
in terms of $F_1(2x_1),\,F_2(2x_2)$ and $C_{\pm}(x_{\pm}),$
solutions of the system (\ref{first}), (\ref{second}):
\ba
\tilde V&=&\frac{1}{2}(C_+' + C_-') + \frac{1}{8}(C_+^2 + C_-^2) +
\frac{1}{4}\biggl( F_2(x_+ -x_-) - F_1(x_+ + x_-)\biggr) ,\nonumber\\
V&=&-\frac{1}{2}(C_+' + C_-') + \frac{1}{8}(C_+^2 + C_-^2) +
\frac{1}{4}\biggl( F_2(x_+ -x_-) - F_1(x_+ + x_-)\biggr) ,\label{potential}\\
B&=&\frac{1}{4}\biggl( C_+ C_- + F_1(x_+ + x_-) + F_2(x_+ - x_-)\biggr) .
\label{functionB}
\ea

The linear character of Eq.(\ref{first}) in $C_{\pm}$ allows to multiply
$C_{\pm}$ by the
imaginary unit keeping $F_{1,2}$ real. This sort
of complexification renders $\tilde V$ of Eq.(\ref{potential})
complex conjugate to $V.$ Thus the intertwining relations (\ref{intertw})
in this case lead automatically to $SUSY-$pseudo-Hermiticity.
For the two-dimensional models the existence of zero modes of $Q^+$ can not
be avoided: actually in the class of models studied in \cite{two}
the equation
for zero modes allows separation of variables ($SUSY-$separation of variables).
Thus these zero modes can be
constructed from normalizable solutions of two one-dimensional equations
of second order (see Section 3 of \cite{two}). In fact, the similarity
relation, which eliminates first order derivatives from the supercharges,
is now unitary.

The next Section will consider the complexification of the partially solvable
(quasi-exactly-solvable) two-dimensional model studied in \cite{two}
(generalized singular Morse potential).
The spectral problem was partially solved by two
methods, one based on the $SUSY-$separation of variables and the second - on the
shape invariance. This model is a natural candidate to elucidate $SUSY-$
pseudo-Hermiticity in two dimensions because it is not amenable to separation of
variables. Furthermore in this model $SUSY-$pseudo-Hermiticity is not
equivalent to pseudo-Hermiticity due to the existence of zero modes of $Q^+$
and also because the spectral problem is not exactly solvable.

We stress that in the class of models (\ref{potential})
the partner Hamiltonians are not \cite{david}-\cite{two}
factorizable in terms of supercharges $Q^{\pm} .$ But there are symmetry
operators of fourth order in derivatives which can be factorized:
\be
R=Q^-Q^+;\quad \tilde R= Q^+Q^- .
\label{sym}
\ee

\section*{\normalsize\bf 3.\quad Complex two-dimensional generalized (singular)
Morse potential.}
\vspace*{0.5cm}
\hspace*{3ex}
The model is defined \cite{two}
in terms of a specific choice for $C_{\pm}$
and $F_{1,2}$ in Eqs.(\ref{potential}) and (\ref{functionB}):
$$
C_+(ia)=4ia\alpha\equiv \hat C_{+}(a);\quad C_-(ia)=
4ia\alpha\cdot\coth \frac{\alpha x_-}{2}\equiv\hat C_-(a);
$$
$$
f_1(x_1)\equiv  \frac{1}{4} F_1(2x_1)=-A\biggl(\exp(-2\alpha x_1) -
2 \exp(-\alpha x_1)\biggr);
$$
$$
f_2(x_2)\equiv  \frac{1}{4} F_2(2x_2)=
+A\biggl(\exp(-2\alpha x_2) - 2 \exp(-\alpha x_2)\biggr);
$$
\ba
\hat V^{\star}(\vec x;a)&\equiv&V^{\star}(\vec x;ia) =
-\alpha^2a(2a+i)\sinh^{-2}\biggl(\frac{\alpha x_-}{2}
\biggr) +\nonumber\\
 &+&A \biggl[\exp(-2\alpha
x_1)-2 \exp(-\alpha x_1) + \exp(-2\alpha x_2)-2 \exp(-\alpha x_2)\biggr];
\nonumber\\
\hat V(\vec x;a)&\equiv&V(\vec x;ia) =
-\alpha^2a(2a-i)\sinh^{-2}\biggl(\frac{\alpha x_-}{2}\biggr) +
\nonumber\\
 &+&A \biggl[\exp(-2\alpha
x_1)-2 \exp(-\alpha x_1) + \exp(-2\alpha x_2)-2 \exp(-\alpha x_2)\biggr],
\label{morse}
\ea
where $A$ is an arbitrary positive constant, and $a$ is a real parameter.
Below we will use for all operators and functions the "hat" notation
following the definitions above. We stress that {\bf only} for {\bf real}
values of the parameter $a$ the model described above satisfies
$SUSY-$pseudo-Hermiticity.

Within this complexification the supercharges $\hat Q^+(a)$  are Hermitian
because $\hat C_{\pm}=\hat C_{\pm}(x_{\pm})$ in Eq.(\ref{ourq}) commute
with $\partial_{\mp}.$ In contrast, the supercharges $Q^-(a)$ for
$a \in \mathbf{R}$ are Hermitian conjugate to $Q^+(a),$ but after
the complexification $a\rightarrow ia$ they are related by
complex conjugation: $\hat Q^-(a)=(\hat Q^+(a))^{\star}:$
\be
\hat Q^{\pm}(a) = 4\partial_+\partial_- \pm 4ia\alpha\partial_-
\pm 4ia\alpha\coth(\frac{\alpha x_-}{2})\partial_+ + \hat B(a).
\label{hatq}
\ee

The Hamiltonian has no definite $PT-$symmetry, but
has a $x_--$reflection symmetry $x_1 \leftrightarrow x_2$
in coordinate space (permutation symmetry). The supercharges (\ref{hatq})
are odd. Therefore this model has vanishing diagonal matrix
elements in (\ref{spectrum}).

In addition, the Hamiltonian has
a discrete symmetry (involution):
\ba
\hat V(\vec x;a)&=& \hat V(\vec x;-a+\frac{i}{2});
\label{symm}\\
\hat V(\vec x;a)&=& \hat V^{\star}(\vec x;-a).
\label{symmm}
\ea

\subsection*{\normalsize\bf  3.1.\quad The method of $SUSY-$separation of variables.}
\hspace*{3ex}
In order to apply the method \cite{two},
one has to separate variables in $\hat Q^{\pm}$
Eq.(\ref{hatq}). This can be achieved by the transformation, which is unitary
for $b \in \mathbf{R} :$
\ba
\hat U(\vec x;b) \equiv \exp \biggl(-ib\alpha( x_+ +
\int\coth(\frac{\alpha x_-}{2})dx_-\biggr) =
\biggl(\frac{\alpha}{\sqrt{A}}\cdot\frac{\xi_1\xi_2}{|\xi_2 -\xi_1|}\biggr)
^{2ib};
\label{U}\\
\hat Q^{-}(0)\equiv \hat U(\vec x;b)\hat Q^{-}(b)\hat U^{-1}(\vec x;b) =
\partial_1^2 -\partial_2^2 +\frac{1}{4}(F_1(2x_1)+F_2(2x_2)),
\label{UU}
\ea
where
$$
\xi_i \equiv \frac{2\sqrt{A}}{\alpha} \exp(-\alpha x_i);\quad i=1,2 .
$$
The zero modes of $\hat Q^+$ can be parametrized as
\footnote{As a consequence of (\ref{UU}), one can derive
$\hat Q^+(0) =(\hat Q^-(0))^{\star}
= \hat U^{\star}(b)\hat Q^+(b)(\hat U^{-1}(b))^{\star}.$}:
\ba
\hat \Omega_n(\vec x;a) &=& \hat U(\vec x;a)) \hat\omega_n(\vec
x;a);\label{Omega}\\
\hat\omega_n(\vec x) &=& \exp(-\frac{\xi_1+\xi_2}{2}) (\xi_1\xi_2)^{s_n}
F(-n, 2s_n +1; \xi_1) F(-n, 2s_n +1; \xi_2),
\label{norm}
\ea
where $F(-n, 2s_n +1; \xi) $ is the standard degenerate (confluent)
hypergeometric function, reducing to a polynomial for integer $n,$ and
\be
s_n=\frac{\sqrt{A}}{\alpha}-n-\frac{1}{2} > 0. \label{si}
\ee

Normalizable eigenfunctions $\hat \Psi_{E_k} (\vec x;a)$ of the
Hamiltonian $\hat H(\vec x;a)$ can be obtained
by linear superposition of zero modes (\ref{Omega}) according to \cite{two},
and their eigenvalues (after complexification $a\rightarrow ia$) read:
\be
\hat E_k(a) = -4ia\alpha^2s_k+4a^2\alpha^2+2\epsilon_k;\quad
\epsilon_k\equiv
-A\biggl[1-\frac{\alpha}{\sqrt{A}}(k+\frac{1}{2})\biggr]^2 < 0.
\label{energy}
\ee
Since the operator $\hat U(\vec x;a)$ in (\ref{Omega}) is unitary,
the condition
for normalizability of eigenfunctions $\hat\Psi_{E_k}(\vec x;a)$ now
does not depend on the parameter $a$ and is expressed by the inequality
(\ref{si}): $s_n>0 .$
The number of normalizable zero modes $\hat\Psi_{E_k}$ is also
determined by this inequality.

Apparently, the energies (\ref{energy}) have nonzero imaginary part
but we remind that (\ref{spectrum}) is trivially satisfied by the vanishing
of the matrix elements $<\hat\Psi_{E_m}|\hat Q^+\hat\Psi_{E_n}>$,
since we deal with zero modes of $\hat Q^+.$

In order to find examples of complex conjugate energies we have to explore
states outside the linear space of zero modes of $\hat Q^+.$
Following the procedure of \cite{two}, we construct three
eigenfunctions
\be
\hat\Phi^{(i)} (\vec x;a)\equiv \hat\Omega_0(\vec x;a)\cdot\hat
\Theta^{(i)} (\vec x;a),
\label{teta}
\ee
with energies ($\hat E_0(a)$ given in (\ref{energy})):
\be
\hat E^{(i)}(a)=\hat E_0(a) + \hat\gamma^{(i)}(a),
\label{gamma}
\ee
and
\ba
&&\hat\Theta^{(1)}(\vec x;a) =
|z_2|^{(4ia+1)};\quad\quad\quad\quad\quad\quad\quad\quad\quad\quad
\hat\gamma^{(1)}(a)= \alpha^2 (2s_0-1)(4ia +1);
\label{a}\\
&&\hat\Theta^{(2)}(\vec x;a) = |z_2|^{(4ia+1)}
\biggl( z_1+\frac{2}{4ia-2s_0+3} \biggr);\quad
\hat\gamma^{(2)}(a)= 4\alpha^2 (s_0-1)(2ia + 1);
\label{bb}\\
&&\hat\Theta^{(3)}(\vec x;a) =
z_1-\frac{2}{4ia+2s_0-1};\quad\quad\quad\quad\quad\quad
\hat\gamma^{(3)}(a)= \alpha^2 \biggl(4ia + 2s_0-1\biggr).
\label{c}
\ea
Here
$$
z_1 = \frac{1}{\xi_1}+\frac{1}{\xi_2};\quad
z_2 = \frac{1}{\xi_1}-\frac{1}{\xi_2}.
$$
In contrast to the case $a \in \mathbf{R},$ where only
$\Phi^{(3)}(\vec x;a)$ is normalizable, all three eigenfunctions
(\ref{a}) - (\ref{c}) become normalizable after $a\rightarrow ia$
if
\be
s_0 > 2. \label{szero}
\ee

One can argue from Eq.(\ref{ps}) that $\hat Q^+\hat\Phi^{(i)}$
are eigenfunctions of $\hat H^{\dagger}$ with the eigenvalues
$\hat E^{(i)}(a).$ As explained after (\ref{ps}), this means that
$\hat H(a),$ in addition to eigenvalues (\ref{gamma}), has
also the complex conjugate eigenvalues:
\be
\hat H(a)(\hat Q^+\hat\Phi^{(i)}(\vec x;a))^{\star} = (\hat E^{(i)}(a))^{\star}
(\hat Q^+\hat\Phi^{(i)}(\vec x;a))^{\star}.
\label{qphi}
\ee
The condition of normalizability of all three wave functions in (\ref{qphi})
coincides with (\ref{szero}).
Orthogonality of all these wave functions can not be secured in general due
to non-Hermiticity of the Hamiltonian\footnote{Due to $x_--$reflection
considerations one can however easily conclude that
$< \hat Q^-(\hat\Phi^{(i)})^{\star}|\hat\Phi^{(j)}> = 0 .$}, but
a pseudo-orthogonality can be derived in agreement with the formalism
\cite{most}-\cite{most2}
of the biorthogonal expansion with particular components
$\hat\Phi^{(i)}; \,\,
\hat Q^-(\hat\Phi^{(i)})^{\star}$ and $(\hat\Phi^{(i)})^{\star};\,\,
\hat Q^+(\hat\Phi^{(i)})$
\ba
<(\hat\Phi^{(i)})^{\star}|\hat\Phi^{(j)}> &=& 0\quad i\neq j \label{1}\\
<(\hat\Phi^{(i)})^{\star}|\hat Q^-(\hat\Phi^{(j)})^{\star}> &=& 0\quad i,j=1,2,3
\label{2}\\
<\hat Q^+\hat\Phi^{(i)}|\hat Q^-\hat\Phi^{(j)}> &=& 0\quad i\neq j \label{3}
\ea
Equations (\ref{3}) can be derived by taking into account \cite{two}
that wave functions $\hat\Phi^{(i)},\, \hat\Phi^{(j)}$ are
eigenfuctions of the symmetry operator
(\ref{sym}) with different eigenvalues. Equation (\ref{1}) is proportional
to (\ref{3}). Eq.(\ref{2}) follows from $x_--$reflection
symmetry considerations ($\hat\Phi^{(i)}$ are even and $\hat Q^{\pm}$
are odd).

Using the symmetry property (\ref{symm}) one can generate a {\bf new series
of levels}
of $\hat H(a)=\hat H(-a+\frac{i}{2})$ by considering each eigenfunction
(and corresponding eigenvalues)
from (\ref{energy}), (\ref{teta}), (\ref{qphi}) with parameter shift
$a \rightarrow
-a+\frac{i}{2}.$  The partner eigenfunctions with complex conjugated
energies can also be constructed along the same line as (\ref{qphi}).

\subsection*{\normalsize\bf 3.2.\quad The shape invariance method.}
\hspace*{3ex}
Starting from a Schr\"odinger equation with potential of (\ref{morse})
$$
\hat H(a)\hat\phi_{E_n}(\vec x;a)= \hat E_n(a)\hat\phi_{E_n}(\vec x;a)
$$
with $\hat\phi_{E_n}(\vec x;a)$ - arbitrary eigenfunction, taking into
account that $\hat H(a+\frac{i}{2})= \hat H^{\dagger}(a),$
we get:
$$
\hat H^{\dagger}(a)\hat\phi_{E_n}(\vec x;a+\frac{i}{2})=
\hat E_n(a+\frac{i}{2})\hat\phi_{E_n}(\vec x;a+\frac{i}{2}).
$$
>From the intertwining relation:
$$
\hat H(a)\hat Q^-(a)=\hat Q^-(a)\hat H^{\dagger}(a),
$$
we obtain
$$
\hat H(a)\biggl(\hat Q^-(a)\hat\phi_{E_n}(\vec x;a+\frac{i}{2})\biggr)
=\hat E_n(a+\frac{i}{2})
\biggl(\hat Q^-(a)\hat\phi_{E_n}(\vec x;a+\frac{i}{2})\biggr)
$$
Thereby, we are at the first step \cite{shape}
of a "shape invariance
chain" of wave functions and eigenvalues for
a complex value of the parameter. Notice
that the definition of potentials now differs from that in \cite{two}
by a constant shift $4\alpha^2a^2.$ This leads to a vanishing of
${\cal R}(a) .$ In addition, we remark that our construction will contain
complex values for the parameters in wave functions,
in operators etc only in intermediate steps,
but the parameter $a$ will always be kept real.

We thus can construct an additional class of levels starting
from $\hat\phi_{E_n},$ an eigenstate of the kind
$\hat\Psi_{E_k};\,\,
\hat\Phi^{(i)};\,\, \hat Q^-(\hat\Phi^{(i)})^{\star}\,\,$ etc
(see (\ref{energy}), (\ref{teta}), (\ref{qphi})).
These states and their complex conjugated are
additional (particular) components of the biorthogonal basis and will fulfill
equations similar to (\ref{1}), (\ref{2}), (\ref{3}).

Iterating this procedure, one generates the shape invariance chain:
\be
\hat\Sigma_{n}^k(\vec x; a) \equiv
\biggl[ \hat Q^-(a)\hat Q^-(a+\frac{i}{2})\hat Q^-(a+i)...
\hat Q^-(a+\frac{i(k-1)}{2})
\hat\phi_{E_n}(\vec x; a+\frac{ik}{2}) \biggr] ,
\label{sigma}
\ee
associated to the energy
$$
\hat E_n^k(a) \equiv \hat E_n(a+\frac{ik}{2}).
$$

In particular, for $\hat\phi_{E_n}$ - linear combination
of zero modes of $\hat Q^+$ (see (\ref{energy}))
\be
\hat\phi_{E_n}(\vec x; a+\frac{ik}{2}) =
\hat\Psi_{E_n}(\vec x; a+\frac{ik}{2})
\label{phipsi}
\ee
the eigenvalues are:
$$
\hat E_n^k(a)= -\alpha^2(4ia(s_n-k)+(s_n-k)^2+s_n^2-4a^2).
$$
In order to investigate the normalizability of these eigenfunctions,
it is crucial to study their behaviour in  $(\xi_1, \xi_2)$
plane following
different paths. As already discussed in Section 4.3. of
\cite{two} for the case $a \in \mathbf{R}$, the relevant
singularities should occur for the origin,
for $\xi_2 \rightarrow 0$ and for $\xi_1\rightarrow \xi_2 .$
By using (\ref{Omega}), (\ref{norm}) and (\ref{UU}) for supercharges and wave
functions, one can study the suitable critical limits in the
following representation for the norm of (\ref{sigma}):
\be
\Vert\hat\Sigma_{n}^k(a)\Vert \sim \Vert\hat U^{-1}(a)\,
\underbrace{\hat Q^-(0)\hat U(-\frac{i}{2})
\hat Q^-(0)\hat U(-\frac{i}{2}) ...
\hat Q^-(0) \hat U(-\frac{i}{2})}_{k \,\, times}
\, \hat U(2a+ik)\hat\omega_n\Vert , \nonumber
\ee
where the $\vec x$ dependence has been dropped for conciseness.
Normalizability can be established for
\be
s_n \equiv \frac{\sqrt{A}}{\alpha} - \frac{1}{2} - n > k ,
\label{last}
\ee
which is just the normalizability condition for (\ref{phipsi}).
In other words the repeated application of $\hat Q^-$ does not restrict
the relevant region of normalizability.

>From the same condition one can estimate the number of normalizable states
generated by successive applications of $\hat Q^-$ operators in (\ref{sigma}).
It depends only on the value of the integer part:
$N\equiv [\frac{\sqrt{A}}{\alpha} - \frac{1}{2}]$ and does not depend on
$a.$ For example, in the case
of non-integer $\frac{\sqrt{A}}{\alpha} - \frac{1}{2}$ the total number of
states can be estimated to be $N(N+1)/2,$ in other words typically
$A/(2\alpha^2)$ for large values of $N.$

As a final remark, we would like to mention that analogous results for shape
invariance chains and their normalizability can be obtained for the model
\cite{two} (before complexification), though they were not explicitly
discussed. In that case $Q^{\pm}$ are
interrelated by Hermitian conjugation, and the calculation of the norm of the
chain can be performed by an explicit introduction of the symmetry operator
$\tilde R = Q^+Q^-,$ provided the arguments match. After this
it is clear that results equivalent to (\ref{last}) hold.

Our main results can be summarized as follows.
In the context of the notion of $SUSY-$pseudo-Hermiticity
two methods (already studied \cite{two} for
Hermitian models) - $SUSY-$separation of variables
and two-dimensional shape invariance - were used to build
explicitly a set of eigenvalues and eigenfunctions for the complex
two-dimensional  singular Morse potential. This part of the spectrum
includes complex conjugated energy pairs and in addition non-paired
complex energies for states - linear superpositions of zero
modes of the intertwining operators.
In contradistinction to one-dimensional models, pseudo-Hermiticity and
$SUSY$-pseudo-Hermiticity are not equivalent for two-dimensional models
just due to the nontrivial role of the zero modes of the supercharges.
For two-dimensional scalar models the intertwining relations can only be
solved for second order supercharges, which a priori may have infinite
number of zero modes. Therefore in the two-dimensional case the intertwined
Hamiltonians are not anymore factorizable (compare with the factorizability
in pseudo-supersymmetry of \cite{mostps}), but  factorizable symmetry
operators (\ref{sym}) of fourth order in derivatives exist.

In the class of models  considered in \cite{two} and here
these symmetry operators cannot be
expressed in terms of  the Hamiltonians and in general signal their
integrability and
possible degeneracy of their spectra, though no direct evidence of such
degeneracy was found in the solved part of the spectrum.
We remind that in \cite{two} the  selected
energy eigenfunctions which were explored  were simultaneously
eigenfunctions of the symmetry operator, so
there was  no indication for degeneracy.

\section*{\normalsize\bf Acknowledgements}
M.V.I. and D.N.N. are indebted to INFN, the University of
Bologna for the support and hospitality.
This work was partially supported by the Russian Foundation for Fundamental
Research (Grant No.02-01-00499).

\end{document}